\documentclass{article}

\usepackage[english]{babel}
\usepackage{amssymb}
\usepackage{amsfonts}
\usepackage{amsmath}
\usepackage{graphicx}
\usepackage{epsfig}
\usepackage{cite}
\makeatletter \@addtoreset{equation}{section} \makeatother

\def\erf#1{(\ref{#1})} % For references to formulas
\newcommand{\cC}{{\cal C}}

  \newcommand{\cL}{{\cal L}}
  
\newcommand{\cO}{{\cal O}}  \newcommand{\cP}{{\cal P}}

  \newcommand{\bZ}{{\mathbf Z}}

  \newcommand{\bq}{{\mathbf q}}
\newcommand{\bx}{{\mathbf x}}

\newcommand{\be}{\begin{equation}} \newcommand{\ee}{\end{equation}}
\newcommand{\bea}{\begin{eqnarray}} \newcommand{\eea}{\end{eqnarray}}
\newcommand{\beann}{\begin{eqnarray*}}  \newcommand{\eeann}{\end{eqnarray*}}
\newcommand{\bfig}{\begin{figure}} \newcommand{\efig}{\end{figure}}
\newcommand{\ba}{\begin{array}} \newcommand{\ea}{\end{array}}
\newcommand{\bcen}{\begin{center}} \newcommand{\ecen}{\end{center}}
\newcommand{\btab}{\begin{tabular}} \newcommand{\etab}{\end{tabular}}
\newcommand{\nn}{\nonumber}
\renewcommand{\Re}{\mathop{\rm Re}}   \renewcommand{\Im}{\mathop{\rm Im}}
\newcommand{\dd}{{\rm d}}
\newcommand{\e}{{\rm e}}

\newtheorem{Proposition}{Proposition}[section]

\newtheorem{Theorem}{Theorem}[section]
\newtheorem{Lemma}{Lemma}[section]
\newtheorem{Corrolary}{Corrolary}[section]

\newcommand{\bp}{\begin{Proposition}}	\newcommand{\ep}{\end{Proposition}}
\newcommand{\bt}{\begin{Theorem}}	\newcommand{\et}{\end{Theorem}}
\newcommand{\bl}{\begin{Lemma}}		\newcommand{\el}{\end{Lemma}}
\newcommand{\bc}{\begin{Corrolary}}	\newcommand{\ec}{\end{Corrolary}}
\begin{document}

\titlepage
\begin{flushright}
IFT-UAM/CSIC-08-45 
\end{flushright}
\begin{center}
{\Large \bf AdS black holes as reflecting cavities} \\[2em]
Irene Amado,${}^a$ Carlos Hoyos,${}^b$ \\  
{\small ${}^a$Instituto de F\'{\i}sica Te\'orica IFT-UAM/CSIC, C-XVI Universidad Aut\'onoma de Madrid\\ 
\small ~\,E-28049 Madrid, Spain\\
\small  ${}^b$Department of Physics,  Swansea University\\  
\small ~\,Swansea, SA2 8PP, UK\\  
\small ~\,E-mail: {\tt Irene.Amado@uam.es, C.H.Badajoz@swansea.ac.uk}}
\date{}
\end{center}
\vspace{2em}
\centerline{ \large \bf Abstract}
{\small We use the identification between null singularities of correlators in the bulk with time singularities in the boundary correlators to study the analytic structure of time-dependent thermal Green functions using the eikonal approximation for classical solutions in the AdS black hole background. We show that the location of singularities in complex time can be understood in terms of null rays bouncing on the boundaries and singularities of the eternal black hole, giving the picture of a `reflecting cavity'. We can then extract the general analytic expression for the asymptotic values of the frequencies of quasinormal modes in large AdS black holes.}

\section{Introduction}

Black holes in asymptotically AdS spaces provide an interesting frame to study the relation between thermal field theories and classical gravity through the AdS/CFT correspondence \cite{Maldacena:1997re,Gubser:1998bc,Witten:1998zw}. The analytic extension of the AdS black hole, the `eternal black hole', has two disconnected boundaries. This has been interpreted as having two independent copies of the field theory each one living in one of the two boundaries \cite{Maldacena:2001kr, Balasubramanian:1998de, Horowitz:1998xk, CarneirodaCunha:2001jf}, based on the original ideas of Israel \cite{Israel:1976ur}. In Schwarzschild time, that is identified with the time in the dual field theory, the second boundary corresponds to the extension to complex values $t\to t-i\beta/2$, where $\beta$ is the inverse of the temperature. This suggests a natural identification in the field theory with the Schwinger-Keldysh formalism\footnote{A description of the Schwinger-Keldysh formalism can be found in Thermal Field Theory books, like \cite{leBellac}.} \cite{Herzog:2002pc}, where the description of the thermal field theory in Lorentzian signature needs to double the degrees of freedom and extend time to complex values. The corresponding Schwinger-Keldysh path starts at some time $t_i$, extends along the real axis to a time $t_f$ and then moves in the imaginary direction to $t_f-i\beta/2$. Then it comes back in the real direction to $t_i-i\beta/2$ and finally it goes to $t_i-i\beta$. The second set of field operators live on the $t-i\beta/2$ piece. 

One of the most interesting results of the correspondence is the relation between the quasinormal mode spectrum that describes the decay of perturbations in black holes and the singularities in the complex frequency plane of two-point correlation functions in the holographic dual \cite{Horowitz:1999jd,Birmingham:2001pj}. Semiclassical computations suggest that both could be simply related to geometric properties of the bulk, in particular to its causal structure \cite{Balasubramanian:1999zv,Louko:2000tp,Kraus:2002iv,Fidkowski:2003nf,Festuccia:2005pi,Hubeny:2006yu}. Very massive fields, corresponding to operators of large conformal dimension, can be studied using a WKB approximation where the field propagates along geodesics. In the black hole background, the space has an analytic extension through the horizon to another asymptotically AdS region. Spacelike geodesics can explore both regions and give information about the thermal state where the field theory is defined. It is also possible to relate different geodesics with the frequency in the field theory, in such a way that in the large frequency limit the geodesic approaches a null ray. The geodesic approximation has been used to compute the asymptotic position of quasinormal modes in the large mass limit, although it has been argued that the results should be generalizable to the case of fields with small mass \cite{Fidkowski:2003nf}.

In this paper we argue that null geodesics in the bulk give useful information about the singularities of the dual correlators even for fields with small mass. In the large frequency limit, that we can associate to the ultraviolet behaviour of the field theory, the classical solutions to the equations of motion can be described in terms of the eikonal approximation. Therefore, the propagation of fields in the bulk is well approximated by null rays and it reduces to a problem of geometric optics in the curved spacetime. In section \ref{sec:mirror} we illustrate this by introducing a mirror in AdS and studying the spectrum of the dual theory. We then proceed in section \ref{sec:eikonal} to find the eikonal approximation for a scalar field in AdS$_{d+1}$ black hole backgrounds and show how the solution bouncing on the singularity can be extended to the asymptotic region behind the horizon. In section \ref{sec:qnms} we explain the asymptotic location of the singularities of field theory correlators and hence the quasinormal mode spectrum of the black hole in simple terms of the geometric shape of the black hole, seen as a reflecting cavity with the asymptotically AdS boundaries and the future and past singularities as walls. Finally, we discuss possible applications of the high frequency-null ray identification and the interpretation in thermal field theories.
%\newpage
\section{AdS with a mirror}\label{sec:mirror}

To illustrate that null geodesics in AdS contain the relevant information about the singularities of the dual two-point correlators let us pick up a very simple example, a scalar field in AdS$_{d+1}$ spacetime. We work with the metric 
\begin{equation}\label{eq:adsmetric}
\dd s^2=\frac{1}{z^2} \left(-\dd t^2+\dd z^2+\dd \bx^2 \right)\,,
\end{equation}
where $z=0$ is the boundary of AdS and we introduce a mirror at a finite value of the radial coordinate $z=z_0$. This translates into Dirichlet boundary conditions for the fields at this surface.

A scalar field $\phi$ with mass $m^2=\Delta (\Delta-d)$ is dual to a scalar operator $\cO$ of conformal dimension $\Delta$. The source $j(t)$ of the operator in the field theory corresponds to a boundary condition for the dual field in the bulk. Consider a spatially homogeneous source localized in time $j(t)=\delta(t)$. Then, the expectation value (vev) of the operator will be given by the two-point correlator $G(t,x)$ as 
\begin{equation}
\langle \cO(t) \rangle \sim \int \dd t' \int \dd\bx \int \dd\bx' G(\bx-\bx',t-t') \delta(t') = V G(t,\bq=0)\,,
\end{equation}
where $V$ is the volume of the space and $\bq$ is the spatial momentum. The singularities of the vev are thus related to singularities of the correlator. In the holographic description, the expectation value is implicit in the asymptotic behaviour of the field, that we can compute using Witten diagrams \cite{Witten:1998zw}. We are interested in the propagation from points $(z=0,t_0,\bx_0)$ at the boundary, to the bulk at $(z,t',\bx')$ and back to the boundary at $(z=0,t,\bx)$. The value of the field at the boundary can be computed using the convolution of two bulk-to-boundary propagators
\begin{equation}
\phi_0(t) \sim \int \dd\bx  \dd\bx_0  \dd\bx'  \dd t_0  \dd t'\int_0^{z_0} \frac{\dd z}{z^{d+1}} \frac{z^{2 \Delta}\; \delta(t_0)}{ \left|z^2+(\bx-\bx')^2-(t-t')^2 \right|^\Delta  \left|z^2+(\bx'-\bx_0)^2-(t'-t_0)^2 \right|^\Delta}
\end{equation}
After integrating over the spatial directions and $t_0$ and introducing Schwinger parameters $w_1$ and $w_2$, we find
\begin{equation}
\phi_0(t) \sim V \int \dd t' \int_0^{z_0}\dd z z^{2 \Delta-d-1} \int_0^\infty \dd w_1  \int_0^\infty \dd w_2\; (w_1w_2)^{\Delta-(d+1)/2} \e^{-w_1\left|z^2-(t-t')^2 \right|} \e^{-w_2 \left|z^2-{t'}^2 \right|}\,.
\end{equation}
The ultraviolet limit corresponds to $w_1\to \infty$, $w_2\to \infty$. The integral is dominated in this case by null trajectories $z=\pm t'$ and $z=\pm(t-t')$. If $t=0$, the two classes of null trajectories become degenerate and there is a singularity. In the presence of a mirror, we can consider a null ray going from the boundary to the mirror and back as part of a single trajectory, so in some heuristic sense the two null trajectories also become degenerate when $t=2 z_0$. In the following we will show that this intuitive picture gives the correct answer by computing explicitly two-point Green functions in the field theory.

The field theory correlator $G(k)$ as a function of the four-momentum $k$ can be computed from the on-shell action for classical solutions of the bulk field $\phi_k(z)$. The result is a boundary term
\begin{equation}\label{eq:boundac}
G(k) \sim \lim_{z\to 0} \sqrt{- g} g^{zz} \phi_k(z) \phi_k'(z)\,.
\end{equation}
For simplicity, we will consider zero spatial momentum and modes with fixed frequency $\phi(t,z)=\e^{-i \omega t} \varphi(z)$. The equations of motion for the field are
\begin{equation}
(\square-m^2)\, \phi = 0 \ \ \Rightarrow \ \ z^{d+1} \partial_z\left( z^{1-d} \varphi'(z)\right)+(z^2\omega^2-\Delta (\Delta-d))\varphi(z) =0\,.
\end{equation}
In order to regularize, we introduce a cutoff at $z=\epsilon$, $\epsilon\to 0$, such that $\varphi(\epsilon)=1$. The solution is given in terms of Bessel functions. Imposing Dirichlet boundary conditions on the field at $z=z_0$ and using $\nu^2=m^2+\frac{d^2}{4}=\left(\Delta-\frac{d}{2}\right)^2$,
\begin{equation}
\varphi(z)=\frac{z^{d/2}\left( Y_\nu(\omega z_0) J_\nu(\omega z) - J_\nu(\omega z_0) Y_\nu(\omega z)\right)}{
\epsilon^{d/2}\left( Y_\nu(\omega z_0) J_\nu(\omega \epsilon) - J_\nu(\omega z_0) Y_\nu(\omega \epsilon)\right)}\,. 
\end{equation}
We then introduce this expression in (\ref{eq:boundac}) and take the limit $\epsilon\to 0$. Up to contact terms, the Green function is given by 
\begin{equation}\label{eq:green1}
G_\Delta(\omega)= c_\Delta \omega^{2 \nu} \frac{ Y_\nu(\omega z_0)}{J_\nu(\omega z_0)}\,.
\end{equation}
If $\nu$ is an integer, there are extra logarithmic terms that cancel the branch cut in $Y_\nu(\omega z_0)$ when $\omega\to 0$. As an example, in AdS$_5$ a massless scalar field gives
\begin{equation}
G_4(\omega)=c_4 \omega^4\left( \frac{\pi Y_2(\omega z_0)}{J_2(\omega z_0)} -\log\left[(\omega z_0)^2\right]\right)\,.
\end{equation}

Apart from a possible branch cut coming from the $\omega^{2 \nu}$ factor, the only singularities of the Green function are poles $\omega_n$ on the real frequency axis, associated to the zeroes of the Bessel function 
\begin{equation}
J_\nu(\omega_n z_0) =0, \ \ n=1,2,3\,\dots
\end{equation}
Since $J_\nu(-x)=(-1)^\nu J_\nu(x)$, the poles are paired $\omega_n$ and $-\omega_n$. 

We are interested in the ultraviolet behaviour of the correlator and how it is related to light-like propagation in the bulk, so we take the $\omega \to \infty$ limit. Then, (\ref{eq:green1}) can be approximated by
\begin{equation}
G_\Delta(\omega)\simeq c_\Delta \omega^{2 \nu} \tan\left[\omega z_0-\left(\nu-\frac{1}{2}\right)\frac{\pi}{2}\right] \,.
\end{equation}
The asymptotic position of the poles is therefore
\begin{equation}\label{eq:asympfreq}
\omega_n z_0=(2n+1)\frac{\pi}{2} + \left(\nu-\frac{1}{2}\right)\frac{\pi}{2}\equiv n\pi +\omega_0 z_0, \ \ n\in \bZ\,.
\end{equation}
To show explicitly the relation with null trajectories is more convenient to look at the time-dependent propagator.
\begin{equation}
G_\Delta(t)=\int_{\cC} \dd \omega \e^{-i \omega t} G_\Delta(\omega)\,.
\end{equation}
For the Feynman propagator, the contour $\cC$ in the complex frequency plane is defined in such a way that it picks up all the positive frequencies for $t\geq 0$ and the negative frequencies for $t<0$, so it passes slightly above the real axis for $\omega>0$ and slightly below for $\omega<0$. Above some frequency $\omega_k$, the position of the poles will be well approximated by the asymptotic expression (\ref{eq:asympfreq}), so the propagator would have a piece coming form the lowest modes plus the contribution from the infinite high-frequency modes. For $t\geq0$, we find
$$
G_{\Delta,F}^+(t)\simeq G_{F,{\rm IR}}^+(t) +2\pi i  c_\Delta (i\partial_t)^{2 \nu} \e^{-i\omega_0 t} \sum_{n=k}^\infty  \e^{-i\pi n t/z_0} =
$$
\begin{equation}
=  G_{F,{\rm IR}}^+(t)+ 2\pi i c_\Delta (i\partial_t)^{2 \nu}  \e^{-i\omega_0 t} \frac{\e^{-i(k-1)\pi t/z_0}}{ \e^{i\pi t/z_0}-1}\,.
\end{equation}
We can understand this expression as being defined using the usual prescription for spacetime-dependent correlators $t\to t-i0^+$.
Strictly speaking, the expression above is well defined only when $2 \nu$ is an integer ($\nu \geq 0$ by unitarity). When this is not the case, there will be a branch cut that must be taken properly into account. This will introduce power-like corrections, but the location of the singularities of the propagator in time is determined by the sum over high-frequency poles. For $t<0$
\begin{equation}
G_{\Delta,F}^-(t)\simeq G_{F,{\rm IR}}^-(t) -  2\pi i c_\Delta (-i\partial_t)^{2 \nu}  \e^{i\omega_0 t} \frac{\e^{-i (k-1)\pi t/z_0}}{\e^{i\pi t/z_0}-1}\,.
\end{equation}

\begin{figure}[!htbp]
\centering
\includegraphics[scale=0.6]{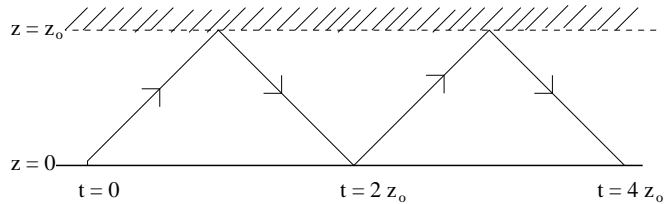} 
\caption{\label{fig:mirrors}\small Null geodesic bouncing on and off from the boundary at $z=0$ and the mirror surface at $z=z_0$.}
\end{figure}

We can see now that the singularities associated to the ultraviolet behaviour of the Feynman propagator appear at regular intervals of time
\begin{equation}
t=2 n z_0, \ \ n\in \bZ\,.
\end{equation}
The identification of these singularities with the singularities on the null trajectories of the bulk propagator leads to the geometric interpretation of a light ray bouncing on and off from the boundary at $z=0$ and the mirror at $z=z_0$, if we take $t=0$ as the `initial point'. The points where the ray reaches the boundary coincide with the singularities in the time-dependent Green function, see figure \ref{fig:mirrors}.

\section{The eikonal approximation in AdS black holes}\label{sec:eikonal}

In black hole backgrounds, the absorption of classical fluctuations of the fields is described by an infinite set of modes with complex frequencies, known as the quasinormal modes. In the holographic dual the frequencies of these modes correspond to singularities of the two-point correlators, that are in last instance responsible for the dissipative behaviour of the thermal theory. The relation between null trajectories ending on the boundary and the high frequency behaviour of the correlators strongly suggests that the analytic continuation of the space behind the horizon can explain the asymptotic location of quasinormal frequencies or equivalently, the singularities of the dual correlators in complex frequency and time.

The eikonal approximation is a high frequency limit where $\omega \gg R$ and $R$ is the typical curvature of the spacetime. This approximation leads to the classical limit of geometric optics in the curved spacetime, similar to ray optics in ordinary electromagnetism. The null trajectories we want to describe start at the AdS boundary and propagate into the interior until they hit the singularity. Presumably it should be possible to extend the geodesic to the asymptotically AdS region behind the horizon by joining it to a null ray that starts at the singularity and continues towards the second boundary. However, the eikonal approximation is expected to fail at the singularity, so it is a matter of concern what is the fate of solutions there. In the following, we construct the classical solutions in the eikonal approximation in the two asymptotically AdS regions and find the matching conditions at the future and past singularities.

In the eikonal approximation, an ansatz for the field is
\begin{equation}\label{eq:ansatz1}
\phi(x) = A(x) e^{i\theta(x)}\,,
\end{equation}
where the eikonal phase $\theta(x)$ is $O(\omega)$ and the amplitude $A(x)$ is $O(1)$. We consider a scalar field with the Klein-Gordon equation of motion
\begin{equation}\label{eq:kleing}
( \square-m^2 ) \, \phi =0 \Rightarrow  \frac{1}{\sqrt{-g}} \partial_\mu \left(\sqrt{-g} g^{\mu\nu} \partial_\nu \phi\right) - m^2 \phi =0 \,.
\end{equation}
Expanding this equation in $\omega$, we find that the leading order gives the eikonal equation
\begin{equation}\label{eq:eikcond}
g^{\mu \nu}\partial_\mu \theta  \partial_\nu \theta =0\,,
\end{equation}
so $k^\mu = \partial^\mu \theta$ is a null vector field. It defines a family of null geodesics tangent to the field and in electromagnetism it can also be identified with the four-momentum of the photons. Notice that the mass is neglected in this approximation, it will appear as a lower energy effect, while the leading behaviour is universal. The next orders give the equations
\begin{eqnarray}\label{eq:eikeqs}
\notag k^\mu \partial_\mu A &  = & -\frac{1}{2} \partial_\mu k^\mu A\,,\\
(\square - m^2)\, A & = &  0\,.
\end{eqnarray}
The first equation describes the evolution of the amplitude along the geodesic, while the last start to take into account subleading effects like the mass.

To work out the eikonal approximation we will use the Rosen coordinate system, that is better adapted to null geodesics. A nice explanation of coordinate systems adapted to the Penrose limit can be found in \cite{blau} and an example of its application to the eikonal expansion in \cite{Hollowood:2008kq}. The AdS$_{d+1}$ black hole metric is ($d\geq 2$) \footnote{We have taken the AdS radius $R=1$, and the coordinates are rescaled as $(t,r,\bx)\to(t/r_{\rm H},r_{\rm H}r,\bx/r_{\rm H})$. Restoring the units, the Hawking temperature of the black hole is $T=dr_{\rm H}/4\pi R^2$.}
\begin{equation}
\dd s^2 = -f(r) \dd t^2+ \frac{\dd r^2}{ f(r)} +r^2\dd \bx^2, \ \ f(r)=r^2-\frac{1}{r^{d-2}}\,.
\end{equation}
We will consider only geodesics at a fixed point in the spatial directions $\bx$, so in terms of the affine parameter $u$ they are determined by two functions $(t(u),r(u))$. These functions can be found by solving the variational problem with Lagrangian
\begin{equation}
\cL =\frac{1}{2}\left(-f(r)\, \dot t^2+\frac{\dot r^2}{f(r)}\right)
\end{equation}
where $\dot t$, $\dot r$ are the first derivatives with respect to $u$. The variation with respect to $t$ introduces a conserved quantity $E$, so that
\begin{equation}
\dot t = \frac{E}{f(r)}\,.
\end{equation}
For null geodesics we can further impose the condition $\cL =0$, so we find
\begin{equation}
\dot r^2 = f(r)^2 \dot t ^2 = E^2\,.
\end{equation}
This also implies that $\ddot{r} =0$. In the Rosen coordinate system we take the affine parameter of the null geodesics $u$ to be one of the coordinates, while we introduce another coordinate $v$ satisfying the null condition $g^{\mu \nu}\partial_\mu v\partial_\nu v =0$ and that corresponds to the Hamilton-Jacobi function of the variational problem. We have several possibilities
\begin{equation}\label{eq:geodesics}
\begin{array}{llll}
1) & E >0, \ \dot r =-E & u=-r & v=-t-\int\frac{\dd r}{f(r)} \\
2) & E >0, \ \dot r = E & u=r & v=-t+\int\frac{\dd r}{f(r)} \\
3) & E <0, \ \dot r =-E & u=r & v=t+\int\frac{\dd r}{f(r)} \\
4) & E <0, \ \dot r = E & u=-r & v=t-\int\frac{\dd r}{f(r)}
\end{array}
\end{equation}
The choices of coordinates 1) and 4) correspond to geodesics starting at the boundary $u=-\infty$ and reaching the singularity at $u=0$. On the other hand, 2) and 3) correspond to geodesics that go from the singularity at $u=0$ to the boundary at $u=\infty$. Geodesics described by 1) and 2) go forwards in time while the ones described by 3) and 4) go backwards. The metric in Rosen coordinates is
\begin{equation}
\dd s^2= 2 \dd u \dd v -f_\pm(u) \dd v^2+u^2 \dd \bx^2\,, \ \ f_\pm(u)=u^2-\frac{(\pm 1)^{d-2}}{u^{d-2}}\,,
\end{equation}
where $f_+(u)=f(u)$ is valid for 2) and 4), while for 1) and 3) we have $f_-(u)$, instead. This distinction is important only when $d$ is odd, since when $d$ is even $f_-(u)=f_+(u)=f(u)$.

We can now write the eikonal ansatz (\ref{eq:ansatz1}) as $\phi(u,v)=A(u)\e^{i\omega v}$ and expand in inverse powers of the frequency
\be\label{eq:eikexpan}
\phi(u,v)=A_0(u)\e^{i\omega v}\left(1+ \frac{A_1(u)}{i \omega}+\frac{A_2(u)}{(i \omega)^2}+\dots\right)\,.
\ee
Plugging back in the equation of motion (\ref{eq:kleing}), we see that the eikonal equation (\ref{eq:eikcond}) is automatically satisfied by our choice of $v$. The next orders in the expansion (\ref{eq:eikeqs}) have the following solution to $O(1/\omega)$
\bea
A_0(u)&=&(-g)^{-1/4}=u^{(1-d)/2}\,, \nn \\
A_1(u)&=&\frac{1}{8}(d^2-1+4m^2)u-(\pm1)^{d-2}\frac{d-1}{8}u^{1-d}\,,
\eea
where the $+$ solution applies for geodesics 2) and 4), while the $-$ is valid for 1) and 3). Then, choosing the correct definition of $v$ and the correct sign in $A_1(u)$, the solution (\ref{eq:eikexpan}) describes the four possible geodesics given by \erf{eq:geodesics}.

This solution is valid as long as $u^{d-1}\omega\gg1$ and $u/\omega\ll1$, so it will stop to be trustable when we get close to the singularity $u=0$ or the boundary $u\to \infty$. There is an extra issue concerning the definition of $v$ in (\ref{eq:geodesics}). The function $f(r)$ has a pole at $r=1$, so in general $v$ will be shifted by a complex value for $r<1$. This could be compensated by defining $v$ differently for $r>1$ and $r<1$ with a compensating constant in this region. The right treatment pass by using Kruskal coordinates, we will analyze this more thoroughly in section \ref{sec:qnms}.

\subsection{Matching of eikonal solutions and bouncing rays}

There are two possible descriptions of a ray bouncing on the boundary, one joining the geodesics associated to 2) with 1) in (\ref{eq:geodesics}) or the time reversed process joining 3) with 4). In terms of the eikonal approximation we should find a matching when we continue the solutions close to the boundary.

At large values of the radial coordinate $r$ the black hole factor can be neglected $f(r)\simeq 1$ and it is better to switch to the coordinate system focused on the boundary (\ref{eq:adsmetric}). In this case, the right identification of eikonal phases will be
$$
\begin{array}{ll}
1) & \e^{i\omega v} \sim \e^{-i\omega (t+r)} \sim \e^{-i\omega (t-z)} \\
2) & \e^{i\omega v} \sim \e^{-i\omega (t-r)} \sim \e^{-i\omega (t+z)} \\
3) & \e^{i\omega v} \sim \e^{i\omega (t+r)} \sim \e^{i\omega (t-z)} \\
4) & \e^{i\omega v} \sim \e^{i\omega (t-r)} \sim \e^{i\omega (t+z)} \\
\end{array}
$$
From section \ref{sec:mirror} we know that the solutions close to the boundary are Bessel functions. To match with the plane wave behaviour of 2), we take the combination that gives the second Hankel function
\begin{equation}
\phi_{(2)} \simeq \e^{-i\omega t} H^{(2)}_\nu(\omega z) \sim \e^{-i\omega (t+z)} , \ \ \omega z \gg 1\,.
\end{equation}
We can treat the bouncing on the boundary as the continuation of this solution to negative values of $z$, $z\to -z$, that we should interpret as a parity transformation of the solutions in the $z$ direction. Under this transformation, the solution changes to the first Hankel function, that shows the right asymptotic behaviour to match with the eikonal solution of 1)
\begin{equation}
\phi_{(1)} \simeq \e^{-i\omega t} H^{(1)}_\nu(\omega z) \sim \e^{-i\omega (t-z)} .
\end{equation}
The time reversed process can be found using the transformation $t\to -t$, so it works in the same way, the initial solution is
\begin{equation}
\phi_{(3)} \simeq \e^{i\omega t} H^{(2)}_\nu(\omega z) \sim \e^{i\omega (t-z)},
\end{equation}
while the reflected one is found again by analytic continuation $z\to -z$
\begin{equation}
\phi_{(4)} \simeq \e^{i\omega t} H^{(1)}_\nu(\omega z) \sim \e^{i\omega (t+z)} .
\end{equation}

We are interested now in rays coming from the boundary that bounce on the future singularity and rays coming from the region behind the horizon that bounce on the past singularity. The expansion of the amplitude (\ref{eq:eikexpan}) fails close to the singularity $u^{d-1} \omega < 1$. However, the ansatz (\ref{eq:ansatz1}) in the form $\phi=A(u) \e^{i\omega v}$ gives still valid solutions to the equations of motion. We can solve the Klein-Gordon equation as a Frobenius expansion, being the general solution of the form
\be\label{eq:singsol}
A(u)=C_1 y_1(u)+C_2 \left[y_2(u)+\log u\,y_1(u)\right]\,,
\ee
where $y_1(u)$ and $y_2(u)$ are series expansions in the $u$ coordinate. For $d\geq2$ we find
\bea
y_1(u)&=&1+\frac{i\omega}{d-1}u^{d-1}-\frac{m^2}{d^2}u^d-\frac{3\omega^2}{4(d-1)^2}u^{2(d-1)} +\dots \nn \\
y_2(u)&=&1+\frac{i\omega}{d-1}u^{d-1}+\frac{d^2-(d-2)m^2}{d^3}u^d+\frac{(4-3d)\omega^2}{4(d-1)^3}u^{2(d-1)}+\dots
\eea
To describe the bouncing on the singularity we need to extend the geodesic associated to 1) and defined for $u<0$ to positive values of the affine parameter. In the original coordinates, $r$ is extended to negative values. The eikonal phase does not change, but the branch cut appearing in (\ref{eq:singsol}) implies that the amplitude will in general pick up a non-trivial phase factor. In order to understand the extended solution from the point of view of an observer in the second boundary, we must change $r\to -r$ keeping $u>0$. The time runs forwards for an observer in the second boundary, so there is no time reversal. The solution is then of the type 2), so we come to a situation analogous to the starting one and we can use the results we have already obtained to describe the bouncing on the second boundary and on the past singularity, that from this perspective looks like a future singularity.

\section{Black holes as reflecting cavities}\label{sec:qnms}

We have shown that in the high frequency limit, the eikonal approximation provides a good description of classical solutions in the bulk, and that the matching conditions at the boundary and singularities are consistent with a limit where a geometric description can be given in terms of null rays bouncing on the boundaries and singularities, so the black hole looks effectively as a box. We have also related null geodesics reaching the boundary with ultraviolet singularities in the time-dependent correlators of the dual field theory and hence with the asymptotic location of singularities in the high frequency limit. We will now proceed to study null geodesics in the black hole in order to extract information about the singularity structure of the correlators in the thermal field theory.

Consider a geodesic starting at the AdS boundary, bouncing on the singularity and reaching the AdS boundary behind the horizon. This corresponds to the analytic extension to positive values of $u$ for the case 1) in (\ref{eq:geodesics}). Notice that $v$ will be shifted in general due to the contribution from $\int \dd r/f(r)$. In order to keep a constant eikonal phase, the value of $t$ has to be shifted as well as part of the analytic continuation. We find then
\begin{equation}
\Delta t = \int_{-\infty}^\infty \frac{\dd u}{f(u)}\,.
\end{equation}
The function $f(u)$ has two poles at the horizon $u=\pm 1$, since for $d$ odd we must switch between $f_-(u)$ and $f_+(u)$. The contour we pick pass above the first pole at $u=-1$ and below the second pole at $u=1$. Then, using the expression
\begin{equation}
\frac{1}{x\mp i \epsilon}=\cP\frac{1}{x}\pm i \pi \delta(x)\,,
\end{equation}
the shift in time on the AdS boundary behind the horizon is
\begin{equation}\label{eq:timeshift}
\Delta t  =  \frac{2\pi}{d} \left( \cot\frac{\pi}{d} -  i \right).
\end{equation}
We have seen that the geodesic coming from the region behind the horizon after bouncing on the past singularity is equivalent to the one we have just described. We can also consider geodesics going backwards in time using the analytic extension of 3), as well as a contour that surrounds the poles at the horizon in the opposite way. We can then identify the complex values of time where the geodesics hit one of the AdS boundaries with the location of singularities of field theory correlators in the complex time plane. The time coordinate is given in dimensionless units, in order to restore the temperature dependence $\beta=1/T$, we must divide by the radius of the horizon $r_H = 4 \pi/d \beta$. For the lowest dimensional AdS$_{d+1}$ spaces, the singularities $t_n$, $n\in \bZ$ are at
\begin{equation}\label{eq:singads}
\begin{array}{ll}
AdS_{2+1} & t_n= i \frac{n\beta}{2} \\
AdS_{3+1} & t_n =\frac{n\beta}{2}\left( \frac{1}{\sqrt{3}}\pm i\right) \\
AdS_{4+1} & t_n=\frac{n\beta}{2} (1\pm i)
\end{array}
\end{equation}
The value of the imaginary part can be understood in terms of the Wick rotation of the metric to Euclidean time (c.f.~\cite{Maldacena:2001kr}), where it can be seen that the second boundary of AdS corresponds to the antipodal point in the thermal circle, with $\beta$ being the full period. The value of the real part can also be understood in simple geometric terms. For this purpose, the best suited coordinate system are Kruskal coordinates
\begin{equation}
U = \e^{2(t+r_*)}, \ \ V=-\e^{-2(t-r_*)}, \ \ r_*=\int \frac{\dd r }{ f(r)}\,.
\end{equation}
with a metric
\begin{equation}
\dd s^2= \frac{-\dd U \dd V }{4 UV f(UV)}+r^2(UV) \dd \bx^2
\end{equation}
In AdS$_3$, the metric takes the particularly simple form
\begin{equation}
\dd s^2= \frac{-\dd U \dd V}{4 (1+UV)^2}+\left(\frac{1-UV}{1+UV} \right)^2 \dd x^2\,.
\end{equation}
The black hole does not cover the entire $(U,V)$ plane, but it is limited by the asymptotic AdS boundary and by the singularity. They can be deduced from the conditions $r^2(UV)\to \infty$ and $r^2(UV)\to 0$, using the relation
\begin{equation}
UV=-\e^{4 r_*(r)}\,.
\end{equation}
For instance,
\begin{equation}
\begin{array}{lll}
AdS_{d+1} & {\rm singularity} & {\rm boundary} \\
d=2 & UV=1 & UV=-1 \\
d=3 & UV=\e^\frac{2\pi}{3 \sqrt{3}} & UV=-\e^\frac{2\pi}{\sqrt{3}} \\
d=4 & UV=1 & UV=-\e^\pi
\end{array}
\end{equation}
The real value of the position of the singularities $t_n$ in the complex time plane can be found following the path of null geodesics in the Kruskal diagram, and using that $t=\log(-U/V)/4$. The null geodesics bounce on the boundaries and the singularities, giving the picture of a reflecting cavity. Following figure \ref{fig:Kruskal}, the points where the singularities are located are
\begin{equation}
\begin{array}{lccl}
AdS_{d+1} & (U,V):\ t=0 \to\; {\rm singularity} \; \to \; {\rm boundary}  &  \\
d=2 &\; (1,-1) \to (1,1) \to (-1, 1)  & \Rightarrow & \Re t = 0 \\
d=3 &  (\e^\frac{\pi }{\sqrt{3}}, - \e^\frac{\pi}{\sqrt{3}}) \to ( \e^\frac{\pi}{\sqrt{3}}, \e^{-\frac{\pi}{3 \sqrt{3}}}) \to (-\e^\frac{7\pi}{\sqrt{3}} , \e^{-\frac{\pi}{3 \sqrt{3}}}) & \Rightarrow & \Re t = \frac{2\pi}{3 \sqrt{3}} \\
d=4 &  (\e^\frac{\pi}{2}, -\e^\frac{\pi}{2}) \to (\e^\frac{\pi}{2},  \e^{-\frac{\pi}{2}}) \to (-\e^\frac{3\pi}{ 2},\e^{-\frac{\pi}{2}}) & \Rightarrow & \Re t =\frac{\pi}{2} 
\end{array}
\end{equation}
In agreement with eq. (\ref{eq:timeshift}). From this formula we can also do a Fourier transformation to deduce the asymptotic quasinormal frequencies, up to a shift that depends on the mass of the field considered. The general formula is
\begin{equation}\label{eq:qnms}
\omega_n \simeq 4\pi T \left(\cot\frac{\pi}{d} \pm i\right)\sin^2\frac{\pi}{d}\; n\,,
\end{equation}
that coincides with the results found in \cite{Cardoso:2004up, Natario:2004jd} using a WKB approximation to the equations of motion in the AdS black hole backgrounds.

\begin{figure}[!htbp]
\centering
{\small
\includegraphics[scale=0.5]{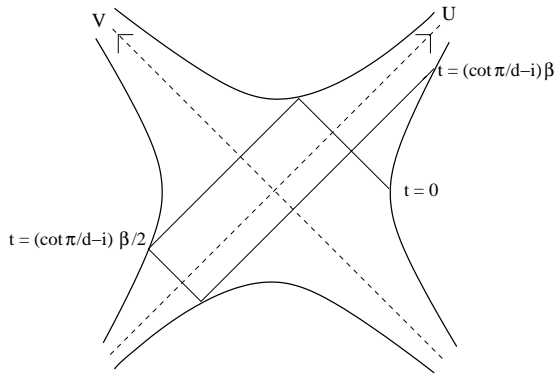} 
\caption{\label{fig:Kruskal}\small  Kruskal diagram of a eternal black hole in AdS space. The light solid line represents a null geodesic starting on the boundary at $t=0$ bouncing on the future singularity, on the second AdS boundary, on the past singularity and back to the boundary at $t=(\cot\frac{\pi}{d}-i)\beta$.}
}
\end{figure}

\section{Discussion}

Using the eikonal approximation, we have seen that the asymptotic formula for the quasinormal spectrum of fields with large mass in AdS black holes is also valid for fields with small mass in the large frequency limit. The analytic formula derived for the the asymptotic value of quasinormal frequencies (\ref{eq:qnms}), agrees with the expressions found in \cite{Cardoso:2004up, Natario:2004jd} using a WKB analysis, and with expressions derived for fields with large mass \cite{Fidkowski:2003nf,Festuccia:2005pi,Hubeny:2006yu} and for fields with small mass in AdS$_{2+1}$ $\omega_n \simeq 4\pi i T  n$ \cite{Birmingham:2001pj}, the numerical results in AdS$_{3+1}$ $\omega_n \simeq (\sqrt{3}\pm 3 i ) \pi T n =  \frac{3}{4}(\sqrt{3}\pm 3 i ) n\, r_{\rm H}$ \cite{Cardoso:2003cj} and AdS$_{4+1}$ $\omega_n\simeq  2\pi i T  n(1\pm i)$ \cite{Horowitz:1999jd}. 

An interesting question is whether a geometric analysis of null geodesics in the analytic extension of the black hole could be applied to more general cases, like charged or rotating black holes or black holes with different asymptotics. A simple example is the topological AdS$_{4+1}$ black hole of \cite{Balasubramanian:2005bg}\footnote{We would like to thank P.~Kumar for pointing out this example to us.}, where the $(t,r)$ part of the metric is the same as AdS$_{2+1}$, so the zero-momentum quasinormal spectrum will have the same asymptotic behaviour. In the dual theory, the correlators are then similar to a free theory, in contrast with the non-extremal black hole. This could be related to the fact that the topological black hole corresponds to D3 branes in the Milne universe, so non-renormalization theorems could still hold. The non-extremal black hole, on the other hand, corresponds to a high temperature phase of the theory.

In this analysis the AdS boundary has played a crucial role since it is there where the value of the quasinormal modes is determined. Other geometries, like asymptotically flat spacetimes, do not have a similar boundary, making more difficult to find a similar prescription. Although in principle there could be a description of extended null geodesics bouncing on the singularities, it is not clear that they can return in the way they do in AdS spacetimes. Nevertheless addressing this issues could give interesting results.

The relation between null geodesics and the high-frequency singularities of the dual correlators also gives a good starting point to solve the inverse holographic problem: how to construct a gravitational background from the field theory. The information contained in the location of the singularities in time is quite topological, it only knows about the causal structure of the space-time. Interesting ideas related to the emergence of the causal structure of the bulk from the field theory can also be found in \cite{Kabat:1999yq,Horowitz:1999gf}. For operators with large conformal dimension, the geodesic approximation could be used to gather more information, see \cite{Porrati:2003na} for instance.

Let us now discuss the holographic interpretation of the black hole computation for Green functions in the thermal field theory. The relation between the Schwinger-Keldysh path and the second boundary of the eternal black hole was pointed out in \cite{Herzog:2002pc}. The analysis shows that there is only one SK path consistent with the prescription of \cite{Son:2002sd} to compute the retarded Green function, corresponding to the symmetric choice
$$
(0,t)\to (t,t-i\beta/2)\to (t-i\beta/2,-i\beta/2)\to (-i\beta/2,-i\beta)\,.
$$
In the thermal field theory, the SK correlator is built introducing insertions of operators at $\Im t=0$ or $\Im t =-i\beta/2$, so it is a matrix with components
\begin{equation}
G_{SK}(t) = \left( \begin{array}{cc} G_F(t) & G^<(t+i\beta/2) \\ G^>(t-i\beta/2) & G_F(t)^\dagger \end{array} \right)\,. \nn
\end{equation}
Where $G^>(t)$, $G^<(t)$ are the Wightman Green functions and $G_F(t)$ is the time-ordered (Feynman) correlator. The short-time singularity of Green functions, together with the KMS relation\footnote{Which can be derived from the periodicity in imaginary time of Euclidean correlators.}
\begin{equation}
G^>(t) = G^<(t+i\beta)\,, \nn
\end{equation}
imply that the SK correlator should have singularities on the time imaginary axis. Since for $d>2$ we find no such singularities, this means that the singularities associated to null rays should correspond to the commutator, that is the spectral function. This is consistent with the identification of the singularities of frequency-dependent correlators with quasinormal modes.

The singularities of frequency-dependent thermal correlators computed from holography turn out to be poles, indicating that there is always an exponential decay with time. For fields with larger conformal dimensions the location of the poles is modified due to the shift of the quasinormal spectrum. Also notice that all this analysis has been made at zero momentum. At non-zero momentum the singularities of the spectral function move in the complex frequency plane, but they remain as poles, in contrast with Green functions at weak coupling where poles open up in branch singularities. Also in \cite{Hartnoll:2005ju} it was argued that higher curvature corrections in the bulk, that correspond to quantum corrections in the field theory, do not change the nature of the singularities. On the other hand the analysis of \cite{Kovtun:2003vj} shows that subleading corrections in the large-$N$ expansion can introduce power-like tails in time. These corrections correspond to quantum corrections in the bulk, so it would be interesting to analyze Green functions in the black hole background beyond the classical approximation.

\paragraph{\large Acknowledgments:} We want to thank T.~Hollowood, P.~Kumar, K.~Landsteiner, C.~Nu\~nez, G.~Shore and D.T.~Son for useful comments and discussions. I.~Amado also wants to thank the hospitality of the Physics Department of Swansea University.  I.\,A. is supported by grant BES-2007-16830 and by the Plan Nacional de Altas Energ\'{\i}as FPA-2006-05485, FPA-2006-05423 and EC Commission under grant MRTN-CT-2004-005104.

%%%%%%%%%%%%%%%%%%%%%%%%%%%%%%%%%%%%%%
%%%%%%%%%% THE BIBLIOGRAPHY %%%%%%%%%%
%%%%%%%%%%%%%%%%%%%%%%%%%%%%%%%%%%%%%%

\end{document}